# Technical Review of Four Different Quantum Systems: Comparative Analysis of Quantum Correlation, Signal-to-Noise Ratio, and Fidelity

Ahmad Salmanogli[1] and Vahid Sharif Sirat[2]

[1]Ankara Yildirim Beyazit University, Engineering Faculty, Electrical and Electronic Department, Ankara, Turkey

[2]Iran University of Science and Technology, Tehran, Iran

**Abstract:** This technical review examines the different methods and approaches have been used to create microwave modes of quantum correlation. Specifically, we consider the electro-opto-mechanical, optoelectronics, 4-coupled qubits, and InP HEMT coupled with two external oscillator methods, and evaluate their effectiveness for quantum applications. Since these systems are open quantum systems, they interact with their own environment medium and thermal bath. To ensure an accurate comparison, we analyzed all of the systems using the same criteria. Thus, at first all systems are introduced briefly, then the total Hamiltonian is theoretically derived, and finally, the system dynamics are analogously analyzed using the Lindblad master equation. We then calculate the quantum correlation between cavity modes, signal-to-noise ratio, and fidelity for each system to evaluate their performance. The results show that the strength and nature of the calculated quantities vary among the systems. An interesting result is the emergence of mixing behavior in the quantum correlation and signal-to-noise ratio for systems that use different cavities. It also identified a significant similarity between the 4-coupled qubits and InP HEMT coupled with external oscillators methods, where an avoided-level crossing occurs in the quantum correlation. Additionally, the study results reveal a high consistency between the signal-to-noise ratio and classical discord.

**Introduction:**

Quantum correlation is a critical quantity used in various quantum applications such as quantum information [1-7], quantum sensors [8-11], and quantum radar [12-14]. In most applications, such as quantum computing and quantum sensing, quantum correlation is created between microwave modes. However, it is fragile and easily destroyed due to thermal noise in the quantum system caused by its interaction with the reservoir modes [5-7]. To prevent this, the quantum system must be operated at sub-cryogenic temperatures of around 10 mK to suppress the thermally excited photons strongly [1-7]. Different methods have been applied to generate microwave photons quantum correlation or quantum entanglement, including electro-opto-mechanical (OC_MR_MC) system [10, 15-16], optoelectronic (OC_PD_MC) [17], coupled qubit [8, 18], and InP HEMT coupling with the two external oscillators [19-

22]. The OC_MR_MC system, shown in Fig.1, is a typical system used to generate entangled photons. It consists of an optical cavity (OC), microresonator (MR), and microwave cavity (MC) coupled to each other. The subsystems are coupled in a way that generates the quantum correlation between cavity modes. However, it has been found that more thermally excited photons can be generated due to the MR operating frequency [10]. This issue will be explained in the next section, where the OC_MR_MC system is analyzed using quantum theory.

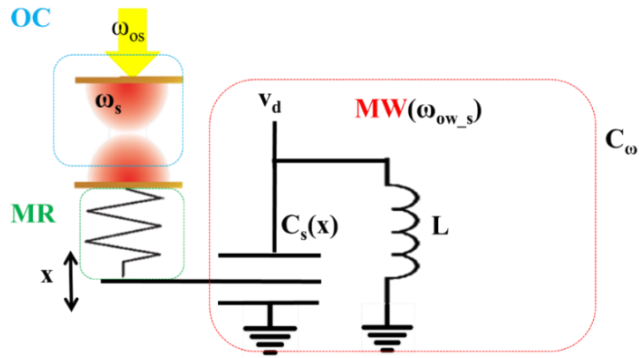

Fig. 1 Electro-opto-mechanical converter; coupling subsystems contain OC, MC, and MR [10].

Due to the arising of some problems in the OC_MR_MC systems, a new quantum system called optoelectronic based converter (OC_PD_MC) has been already introduced by the author [17]; in the design mentioned, the MR subsystem replaced by an optoelectronic component schematically shown in Fig. 2. The MR is removed from the converter, so the optoelectronic component is employed to establish coupling between MC and OC modes. This led to high frequency operation due to the strong decrease in thermally excited photons. In this system, the OC modes are initially coupled with the photodetector (PD), and then detected photons are coupled with MC. The operation of the system is discussed in detail in the next section [17]. Another alternative method for efficient generation of microwave quantum correlation offers operation of qubits at sub-cryogenic temperature approximately around 10 mK; the quantum correlation can be achieved at a frequency of around 5.2 GHz [8, 18]. Qubits have unique properties which make them as a popular choice for consideration in quantum computation [23-27]. Quantum computation exploits qubits' quantum superposition and entanglement properties, which are responsible for their significant speed and efficiency advantages over classical computers [23]. In the case of quantum computer, the entanglement process between two qubits is created through the non-linear properties of the Josephson Junction (JJ) of a qubit, and the coupling effect of the qubits [23-27]. The qubits' ground state transits to other qubits' multiple excited states due to the entanglement process. A seminal work in this area [23] has theoretically and experimentally demonstrated that the entangled states give rise to two transitions out of the ground state and lead to an avoided level-crossing phenomenon in the spectroscopy of the coupled qubits. This behavior occurs due to the coupling of the RF incident wave, which induces transitions out of the ground state. In order to better understanding and comparing the quantum properties of the qubits with that of other mentioned quantum systems, a quantum circuit comprising four coupled qubits (called 4-Qubits in this article) has been designed and analyzed using quantum mechanics method [18]. This system and its operation will be discussed in the next section.

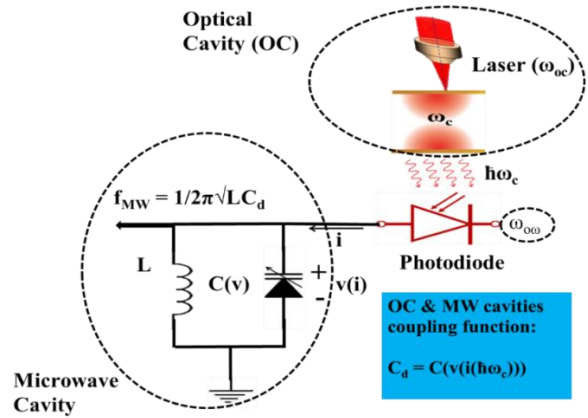

Fig. 2 Optoelectronic converter contains OC, optoelectronic device, MC [17].

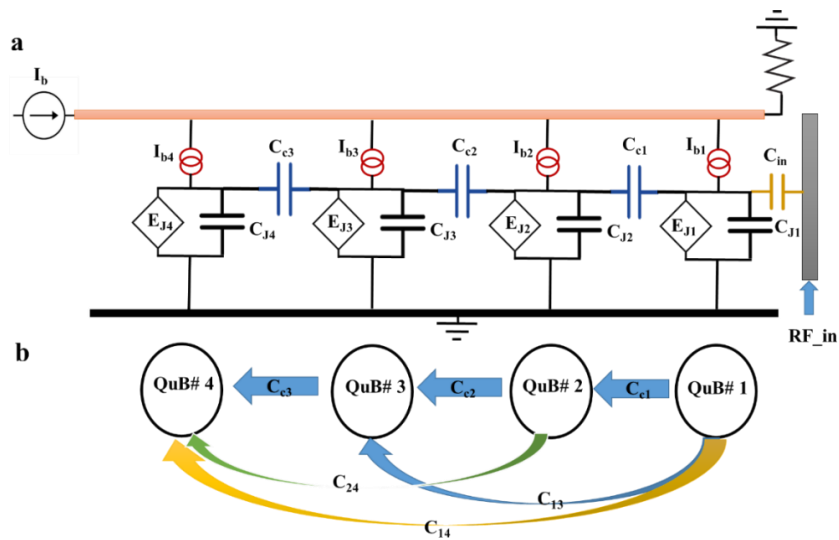

Fig. 3 Schematic of the Coupling of 4 qubits to each other, b) schematic of the capacitive connection of the qubits to each other [18].

Another interesting method for generating microwave photons of quantum correlation introduces electronic circuits operating at cryogenic temperatures (Fig. 4) [19]. This circuit operates as a low noise amplifier (LNA), focusing on minimizing the noise figure. The circuit in Fig. 4, comprises InP HEMT as a nonlinear active element, an input and output matching network for impedance matching, and a DC stabilization circuit. The inset figures show the generic equivalent circuits of the input and output matching network [19]. InP HEMT is biased via $V_g$ and $V_d$; also, the small signal RF input is applied through an input capacitor ($C_{in}$). $L_{g1}$ and $L_{d1}$ are critical components in the stabilization circuit. The transistor nonlinear equivalent circuit is also included as another inset figure, which displays all factors that could influence on the InP HEMT's operation at cryogenic temperatures [19]. This circuit has been designed for quantum applications, where the number of microwave photons is kept at a low level, since thermally excited photons can negatively affect the low photon quantum applications [20]. Therefore, the noise generated by all resistances is considered. The circuit's critical component is $i_{ds}$, a dependent current source controlled by the voltage. The current's value is manipulated by $g_m$, intrinsic transconductance; also, it is additionally influenced by $g_{m2}$ and $g_{m3}$, referred to as the second and third nonlinearity factors. In summary, the quantum correlation generation between microwave photons using electronic circuits operating at cryogenic temperatures can be considered as an effective approach. The Hamiltonian of the derived circuit and all parameters factors may affect microwave quantum correlation are discussed in detail in the following section [19-21].

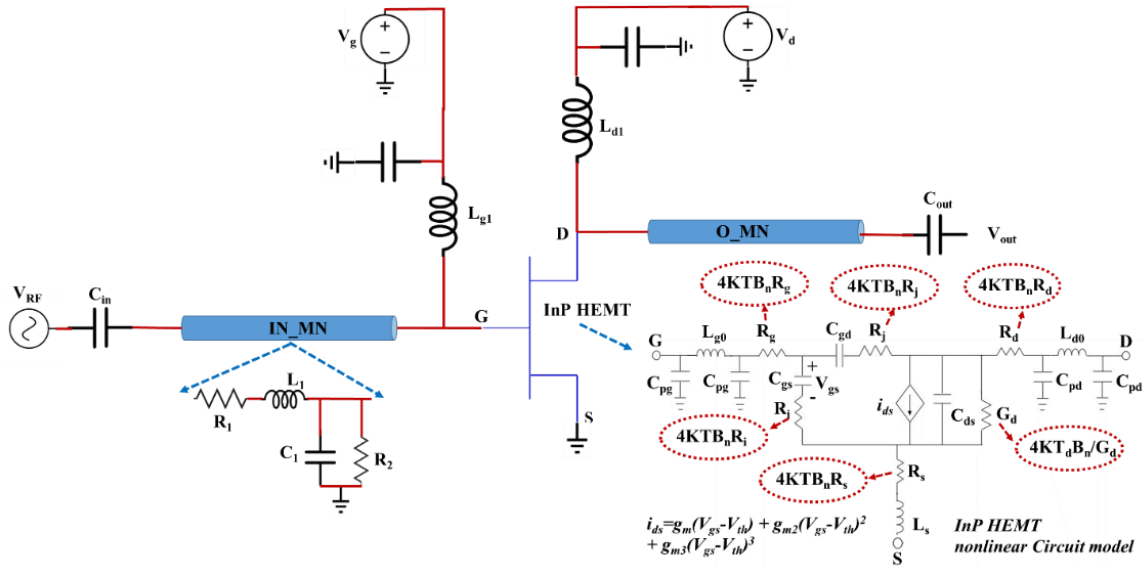

Fig. 4 the circuit schematic containing the input and output matching networks, stability network, and active elements (InP HEMT) and its internal circuit [19].

This study compares the performance of four selected different quantum systems, namely OC_MR_MC, OC_PD_MC, InP HEMT coupling to external oscillators, and 4-Qubits discussed shortly above, in terms of their microwave quantum correlation, classical correlation, fidelity, and signal-to-noise ratio (SNR). We aim to provide a comprehensive understanding of the behavior and restrictions of these systems, which can help reader to select the most suitable system for the desired application.

**Materials and methods:**

*A short review of the design of the electro-opto-mechanical (OC_MR_MC)*

In an OC_MR_MC system, which can be considered as a tripartite quantum system, the main process initiates with exciting the OC by a signal that can generate quantized optical modes. The OC output modes are then coupled to the MR via optical pressure, leading to resonating MR with a frequency in the MHz range. The MR resonator oscillation alters the MC capacitance, which make changes the microwave cavity's resonant frequency. The change in the capacity indicates that MR couples to MC. It has been shown that some critical parameters can be manipulated for system engineering and improving the modes entanglement. Coupling factor among cavities, cavities decaying rate, frequency resonance of the cavities, and OC and MC driving field, are the mentioned parameters have been theoretically derived and presented in [10, 15-16]. By manipulating these parameters, one can establish the non-classical correlation between the cavities' modes. Since the modes of OC_MR_MC cavities are coupled to each other, the quantum entanglement between OC and MC modes can be created by engineering the converter's parameters. However, the mode entanglement generated can be easily destroyed due to the cavities' coupling to the reservoir modes. Therefore, the term quantum entanglement is used in the OC_MR_MC system where operated at a sub-cryogenic temperature (<10 mK) rather than quantum correlation. The converter in Fig.1 consists of three different parts: OC, MR, and MC, which are coupled to each other. The usual way to study the system is to prepare the contributed systems' Lagrangian [28-29]. In the following, the total quantum Hamiltonian is defined to clarify the system's dynamic equation and is presented as [10]:

$$H_{OC} = \hbar \omega_c a_c^+ a_c$$

$$H_{MR} = \frac{\hbar \omega_m}{2}(P_x^2 + q_x^2)$$

$$H_{MC} = \hbar \omega_\omega c_\omega^+ c_\omega - jV_d C_d \sqrt{\frac{\hbar \omega_\omega}{2C_t}}(c_\omega - c_\omega^+)$$

$$H_{OC-MR} = \hbar \sqrt{\frac{\alpha_c^2 \omega_m}{2\varepsilon_0 \omega_c m}}(a_c^+ + a_c)P_x \qquad (1)$$

$$H_{MC-MR} = C_p C_t \frac{\hbar \omega_\omega}{2}\sqrt{\frac{\hbar}{\omega_m m}} q_x (c_\omega - c_\omega^+)^2$$

$$H_{OC-drive} = j\hbar E_c (a_c^+ e^{(-j\omega_L t)} - a_c e^{(j\omega_L t)})$$

In Eq. (1), $H_{OC}$, $H_{MR}$, $H_{MC}$, $H_{OC-MR}$, $H_{MC-MR}$, and $H_{OC-drive}$ are the Hamiltonians for the OC, MR, MC cavities, OC-MR interaction, MC-MR interaction, and OC cavity driving, respectively. $(P_x, q_x)$, $(\omega_c, \omega_m, \omega_\omega)$, $\hbar$, j, $V_d$, and $\alpha_c$ are the MR relating quadrature operator, the OC, MR, and MC cavity relating frequencies, reduced Planck constant, imaginary number, MC driving voltage, and coupling coefficients between OC and MR, respectively. $(a_c^+, a_c)$ and $(c_\omega^+, c_\omega)$ represent the creation and annihilation operators for the OC and MC modes, respectively. $E_c$ corresponds to the OC driving field [15-16]. $C_p$ and $C_t$ are calculated using $C_p = C'(x)/[C(x) + C_d]^2$ and $C_t = C_d+C(x_0)$, where $x_0$ represents the equilibrium position of the MR resonator [10]. The second term in $H_{MC}$ describes the external source cavity driving. The dynamic equation of motion for the system is routinely derived using the Heisenberg-Langevin equations [28-29], taking into account the system's interaction with the environment, which includes to effect of noise and damping rates. Therefore, the system equation of motion becomes as [10]:

$$\dot{a}_c = -(i\Delta_c + \kappa_c)a_c - jG_1P_x + E_c + \sqrt{2\kappa_s}a_{in}$$
$$\dot{c}_\omega = -(i\Delta_\omega + \kappa_\omega)c_\omega + j\Delta_{o\omega 1}G_2q_xc_\omega + E_\omega + \sqrt{2\kappa_\omega}c_{in}$$
$$\dot{q}_x = \omega_mP_x + G_1(a_c^+ + a_c)$$
$$\dot{P}_x = -\gamma_mP_x - \omega_mq_x + \Delta_\omega G_2c_\omega^+c_\omega + b_{in}$$
(2)

In Eq. 2, $\kappa_c$, $\gamma_m$, $\kappa_\omega$, $\Delta_c$, and $\Delta_\omega$ correspond to the damping rates and detuning frequencies of OC, MR, and MC, respectively. $G_1 = \sqrt{(\alpha_c^2\omega_m/2\varepsilon_0m\omega_c)}$, $G_2 = C_pC_t\sqrt{(\hbar/m\omega_m)}$, and $E_\omega$ are the coupling strengths between OC and MR, between MR and MC, and the MC driving field, respectively [15-16]. Eq. 2 reveals the interdependence between the coupled cavities, which allows for the identification of critical parameters that influence the cavity modes. Therefore, the quantum correlation between modes can be investigated and evaluated. The data presented parameters are given in Table 1 are utilized to simulate and analyze the system.

Table 1. Data used to model the OC_MR_MC [10, 15-16]

| | |
|---|---|
| $\lambda_c$ (incidence wavelength) | 808 nm |
| $f_m$ (MR resonance frequency) | 10 MHz |
| $f_\omega$ (MW resonance frequency) | 10 GHz |
| $L_c$ (OC length) | 1 mm |
| $L_{MC}$ (MC length) | 10 nm |
| $P_c = P_\omega$ (Cavities driving power) | 30 mW |
| $\Delta_{OC} = \Delta_{MC}$ (detuning frequency) | $\omega_m = 0.01*2\pi f_m$ |
| $\gamma_m$ (MR damping rate) | ~1000 (1/s) |
| $\kappa_c$ (OC damping rate) | $0.08\omega_m$ |
| $\kappa_\omega$ (MW damping rate) | $0.02\omega_m$ |
| m (MR resonator mass) | 20 ng |
| Tem (environment temperature) | 50 mK |

Generally, when a system interacts with the environment, its dynamics are described by the Lindblad master equation [16]. This equation is necessary because a quantum system becomes stochastic during interaction with the environment, so utilizing the Schrödinger equation is no longer satisfactory. The state of the open system is described by the density matrix formalism, which defines the probability distribution of the related quantum state.

In this article, similar to what is done in the literature, similarly to [29-30], it is assumed that for all systems discussed, interacting with the environment, the system is expanded to include the environment. Thus, the expanded system (the system combined with the environment) can be considered a closed quantum system.

The evolution of the closed quantum system is governed by the von Neumann equation, which can be derived as follows [29-30]:

$$\dot{\rho}(t) = \frac{1}{j\hbar}[H_t, \rho_t(t)] \tag{3}$$

where $\rho(t)$ and $H_t$ are the density matrix describing the probability distribution of a quantum state and system total Hamiltonian. Generally, the total Hamiltonian is expressed as $H_t = H_0 + H_{int} + H_{env}$, where the terms, respectively, define free Hamiltonian, interaction Hamiltonian, and environmental Hamiltonian. Here, the critical point is the dynamics of the quantum system that we are interested in; thus, one can perform the partial trace over $H_{env}$ to attain the master equation. The complete form of the master equation is the Lindblad master equation defined as [30]:

$$\dot{\rho}(t) = \frac{1}{j\hbar}[H, \rho(t)] + \frac{1}{2}\sum_n [2C_n\rho(t)C_n^+ - \rho(t)C_n^+C_n - C_n^+C_n\rho(t)] \tag{4}$$

The Lindblad master equation, which describes the dynamics of a quantum system interacting with its environment, is governed by the collapse operator $C_n$. Previous research [31] has analyzed the factors affecting the coupling between a quantum system and its reservoir, including the decay rate generated by the interaction. However, several assumptions are also necessary to derive the master equation in Eq. 3, such as the absence of correlation between the system and environment at t = 0, separability during evolution, and minor changes to the environment's state. To solve the Lindblad master equation, the Qutip toolbox in Python [30] is utilized to determine the time evolution of the system's density matrix and ultimately, the covariance matrix (CM) for the quadrature operators $X_i$ and $Y_i$, where i = 1, 2. These operators are defined for the OC and MC using the formulae $X_i = (a_i + a_i^+)/\sqrt{2}$ and $Y_i = (a_i - a_i^+)/j\sqrt{2}$.

To quantify the quantum correlation of the system, the quantum discord is calculated as a known quantifier [5-7]. The compact forms of quantum discord ($Q_{discord}$) and classical discord ($C_{discord}$) are presented as $D(\rho_{AB}) = h(b) - h(v_-) - h(v_+) + h(\tau + \eta)$ and $C(\rho_{AB}) = h(a) - h(v_-) - h(v_+)$, respectively. Here, $v_\pm$ is the Symplectic eigenvalue of the CM, and $h(x) = (x+0.5)\log_2(x+0.5) - (x-0.5)\log_2(x-0.5)$ [7,13]. The variable "$b$" represents the second cavity output photon number, while $\tau$ and $\eta$ are the quantities defined in terms of the first cavity output photon number "$a$" and the phase-sensitive cross-correlation between the two coupled cavities $d_{o12}$ [5-7]. The last term in quantum discord, $h(\tau + \eta)$, defines the measurement effects on the second cavity. The analysis shows that various factors, such as the second cavity average photon number, the Symplectic eigenvalue of the CM, and the classical correlation effect, influence the quantum correlation between modes. Notably, the phase-sensitive cross-correlation do$_{12}$ plays a critical role in determining quantum discord [8].

*A short review of the design of the optoelectronic converter (OC_PD_MC)*

In this section, we focus on the design of the optoelectronic converter (OC_PD_MC). In this system, the mechanical component operating at low frequency can cause some problems. The low-frequency operation generates many photons due to the thermal effect, which can destroy entanglement. Therefore, to alleviate the problem, it has been proposed to replace the mechanical component with an optoelectronic component [17]. A schematic diagram of this system is shown in Fig. 2. To produce the entanglement, an optoelectronic converter has been theoretically designed, where OC is connected to MC through a Varactor diode (VD), which is excited by a PD. The device operates as follows: OC modes initially excite the PD, and the current flow due to the incident light triggers the VD. The current flow due to the OC coupling with the PD causes a voltage drop across the diode, which creates a functional relationship between the drop voltage and the OC modes. In some literatures [17], the effect of critical quantities, such as the PD coupling factor ($\mu_c$) on the MC, has been extensively studied for engineering the system. The optoelectronic converter has been theoretically analyzed using the canonical quantization method [18], which emphasizes the interaction between the incident and the atom's quantum field, in contrast to the dipole approximation method [29]. To define the dynamics of a quantum system, the total Hamiltonian of the system must be expressed [17], which is:

$$H_{OC} = \frac{\varepsilon_0}{2}(E^2 + \omega_c^2 A^2)$$

$$H_{PD} = \frac{P^2}{2m_{eff}} + \frac{1}{2}m_{eff}\omega_{eg}^2 X^2$$

$$H_{MC} = \frac{Q^2}{2C_0} + \frac{\phi^2}{2L} + \frac{C_d v_d}{C_0}Q \tag{5}$$

$$H_{OC-PD} = \alpha_c \frac{AP}{m_{eff}}$$

$$H_{MC-PD} = \frac{-C'(x)}{C_0^2}\left\{\frac{Q^2}{2} + C_d v_d Q\right\}X$$

In these equations, the (X, P), (A, E), and (Φ, Q) are the PD electron-hole position and momentum operators, OC vector potential and electric field, and the MC phase and charge operators, respectively. The constants $\alpha_c$, $v_d$, C'(x), and $C_d$ represent the OC-MC coupling coefficient, the MC driving field, the variable capacitor, and the capacitor between the MC driving field and cavity, respectively [17]. To represent the quantum Hamiltonian in terms of the raising and lowering operators, it is necessary to determine the conjugate variables of operators (A, X, and Φ) based on the classical approach [29]. The raising and lowering operators are given by [17]:

$$H_{OC} = \hbar\omega_c a_c^+ a_c + j\hbar E_c[a_c^+ e^{(-j\omega_{oc}t)} - a_c e^{(j\omega_{oc}t)}]$$

$$H_{PD} = \frac{\hbar\omega_{eg}}{2}(P_x^2 + q_x^2)$$

$$H_{MC} = \hbar\omega_\omega c_\omega^+ c_\omega - j|v_d|C_d\sqrt{\frac{\hbar\omega_\omega}{2C_0}}[c_\omega e^{(-j\omega_{o\omega}t)} - c_\omega^+ e^{(j\omega_{o\omega}t)}] \quad (6)$$

$$H_{OC-PD} = \hbar\sqrt{\frac{\alpha_c^2 \omega_{eg}}{2\varepsilon_0 m_{eff}\omega_c}}(a_c^+ + a_c)P_x$$

$$H_{MC-PD} = -\frac{\hbar\mu_c\omega_\omega}{2d}\sqrt{\frac{\hbar}{\omega_{eg}m_{eff}}}q_x c_\omega^+ c_\omega$$

where the operators ($a_c^+$, $a_c$), and ($c_\omega^+$, $c_\omega$) are the creation and annihilation operators for the OC and MC, respectively. The operators ($P_x$, $q_x$), $E_c$, $g_{op}$, and d are normalized quadrature operators, driving rate of optical cavity input, and the depletion layer width of VD's capacitor, respectively, and $\mu_c = C'(x)/C_0$. The interaction Hamiltonian between the MC and PD, $H_{MC-PD}$, is defined as part of the overall Hamiltonian. The coupling factor in $H_{MC-PD}$, $g_{o\omega} = (\mu_c\omega_\omega/2d)\times\sqrt{(\hbar/\omega_{eg}m_{eff})}$, is introduced to enable the manipulation of quantum correlation in the system. In Eq. 6, the second term in $H_{MC}$ refers to MC driving, and $g_{op} = \sqrt{(\omega_{eg}\alpha_c^2/2\omega_c\varepsilon_0 m_{eff})}$ defines the OC-PD coupling rate. To calculate $g_{op}$, which is a critical factor in the system, first-order perturbation theory has been employed, as expressed in previous studies [17].

$$g_{op} = \frac{\pi\omega_c}{\varepsilon_0 V_m}\mu^2 g_J(\hbar\omega_{eg})L(\omega_{eg}) \rightarrow \alpha_c = \sqrt{\frac{2\omega_c\varepsilon_0 m_{eff}}{\omega_{eg}}\frac{\pi\omega_c}{\varepsilon_0 V_m}\mu^2 g_J(\hbar\omega_{eg})L(\omega_{eg})} \quad (7)$$

In Eq. 7, $g_J(h\omega_{eg})$, $\mu$, and $L(\omega_{eg})$ are the PD density of state, dipole momentum, and Lorentzian function, respectively. Notably, $\alpha_c$ is strongly dependent on the electrical and optical properties of the PD. Thus, this gives a reasonable degree of freedom for engineering and manipulating the coupling between OC and PD.

Table 2. Data for simulation of OC_PD_MC [17]

| | |
|---|---|
| $\lambda_c$ (incidence wavelength) | 808 nm |
| $f_{PD}$ (optoelectronic resonance frequency) | 1 GHz |
| $f_\omega$ (MC resonance frequency) | 10 GHz |
| $L_c$ (OC length) | 1 mm |
| $P_c = P_\omega$ (Cavities driving power) | 30 mW |
| $\Delta_{OC} = \Delta_{MC}$ (detuning frequency) | $\omega_m = 0.01*2\pi f_m$ |
| $\kappa_c$ (OC damping rate) | $0.08\omega_m$ |
| $\kappa_\omega$ (MC damping rate) | $0.02\omega_m$ |
| $M_{eh}$ (Electron-hole effective mass) | $5.11\times10^{-32}$ g |
| Tem (environment temperature) | 50 mK |

In this system, it is also necessary to add the damping rate and noise effects for dynamic equations because of the system and real medium interaction. Also, using rotating wave approximation (RWA), one can define the detuning frequencies of MC, OC, and PD as $\Delta_\omega = \omega_\omega - \omega_{o\omega}$ and $\Delta_c = \omega_c - \omega_{oc}$, and $\Delta_{eg} = \omega_{eg} - \omega_c$, respectively. Therefore, the dynamic equation of motion of the quantum system discussed [17] is given by:

$$\dot{a}_c = -(j\Delta_c + \kappa_c)\hat{a}_c - jg_{op}\hat{P}_x\hat{a}_c + E_c + \sqrt{2\kappa_c}\,\hat{a}_{in}$$
$$\dot{c}_\omega = -(j\Delta_\omega + \kappa_\omega)\hat{c}_\omega + jg_{\omega p}\hat{q}_x\hat{c}_\omega + E_\omega + \sqrt{2\kappa_\omega}\,\hat{c}_{in}$$
$$\dot{q}_x = \Delta_{eg}\hat{P}_x + g_{op}(\hat{a}_c^+ + \hat{a}_c) \tag{8}$$
$$\dot{P}_x = -\gamma_p\hat{P}_x - \Delta_{eg}\hat{q}_x + g_{\omega p}\hat{c}_\omega^+\hat{c}_\omega + \hat{b}_{in}$$

where $\kappa_c$, $\gamma_p$, and $\kappa_\omega$ are, respectively, the OC, PD, and MC damping rates. Since the cavities interact with the environment, the quantities such as $b_{in}$, $a_{in}$, and $c_{in}$ as the noise sources are defined to apply and model of the environmental effect on cavities. Finally, the Lindblad master equation is applied to govern the dynamics of the presented system in the same way as the OC_MR_MC system. In addition, the data listed in Table.2 are used to simulate and analyze the discussed quantum system.

*A short review of the design of the 4 coupled qubits (4-Qubits):*

In this article, we also discuss a quantum system consisting of four coupled qubits (4-Qubits) that was designed in [18] and is schematically illustrated in Fig. 3a. The system is capacitively coupled to an RF wave through a capacitor ($C_{in}$). Each qubit contains a Josephson junction with associated energy ($E_J$) and a parallel capacitor ($C_J$) for storing the charging energy. All qubits are biased with $I_b = I_c + \Delta I_c$, where $\Delta I_c$ is a ramp function. Fig. 3b shows the direct and indirect coupling of the qubits in the system. From a quantum point of view, the ground state of each qubit can transit to every four excited states [18, 23], meaning that the coupled qubits, such as QuB#1 with QuB#4 should exhibit a correlated state [18]. To analyze the system, it is necessary to derive the total quantum Hamiltonian. Thus the Lagrangian of the quantum circuit is initially derived as [18]:

$$L_{QC} = \frac{C_{J1}}{2}\left(\dot{\Phi}_1\right)^2 + E_{J1}\cos(\frac{2\pi\Phi_1}{\Phi_0}) + \frac{C_{in}}{2}\left(V_{rf}-\dot{\Phi}_1\right)^2 + \frac{C_{c1}}{2}\left(\dot{\Phi}_1-\dot{\Phi}_2\right)^2 + \frac{C_{13}}{2}\left(\dot{\Phi}_1-\dot{\Phi}_3\right)^2 + \frac{C_{14}}{2}\left(\dot{\Phi}_1-\dot{\Phi}_4\right)^2$$
$$+ \frac{C_{J2}}{2}\left(\dot{\Phi}_2\right)^2 + E_{J2}\cos(\frac{2\pi\Phi_2}{\Phi_0}) + \frac{C_{c2}}{2}\left(\dot{\Phi}_2-\dot{\Phi}_3\right)^2 + \frac{C_{24}}{2}\left(\dot{\Phi}_2-\dot{\Phi}_4\right)^2$$
$$+ \frac{C_{J3}}{2}\left(\dot{\Phi}_3\right)^2 + E_{J3}\cos(\frac{2\pi\Phi_3}{\Phi_0}) + \frac{C_{c3}}{2}\left(\dot{\Phi}_3-\dot{\Phi}_4\right)^2 \tag{9}$$
$$+ \frac{C_{J4}}{2}\left(\dot{\Phi}_4\right)^2 + E_{J4}\cos(\frac{2\pi\Phi_4}{\Phi_0})$$

In Eq. 9, $\Phi_i$, $V_{rf}$, $C_{in}$, and $C_{ci}$, $C_{ik}$ (i = 1,2,3,4 and k = 3, 4) represent the qubit flux, RF incident wave amplitude, input capacitor used to couple the incident RF to the quantum circuit, and coupling capacitors between qubits, respectively. The momentum conjugate variable of the fluxes ($Q_i$) is obtained using the Legendre transformation. The canonical quantization method is applied to the classical degree of freedom to obtain $[\Phi_i, Q_i] = j\hbar$, where $j = \sqrt{(-1)}$. The quantum Hamiltonian is then derived as [18]:

$$H_{QC} = \sum_{n=1}^{4} \sqrt{8 E_{cn} E_{Jn}} \left( b_n^+ b_n \right) - \frac{I_{bn}}{\sqrt{2\pi}} \left( \frac{8 E_{cn}}{E_{Jn}} \right)^{0.25} \frac{\Phi_0}{2\pi} \left( b_n^+ + b_n \right) - \frac{E_{cn}}{12} \left( b_n^+ + b_n \right)^4 - j m_n \frac{I_{bn}}{\sqrt{2}} \left( \frac{E_{Jn}}{8 E_{cn}} \right)^{0.25} (2e) V_{rf} \left( b_n - b_n^+ \right) \\ - \sum_{\substack{m=1 \\ m \neq n}}^{4} 4 E_{c\_nm} \left( \frac{E_{Jn}}{8 E_{cn}} \right)^{0.25} \left( \frac{E_{Jm}}{8 E_{cm}} \right)^{0.25} \frac{\left( b_n - b_n^+ \right)\left( b_m - b_m^+ \right)}{2} \tag{10}$$

where $\Phi_0$ is the flux quantum, and coefficients such as $m_n$, $C_n$, and $C_{nm}$ are defined in detail in [18]. In this equation, some definitions are used, such as $\hbar \Omega_n = \sqrt{8(E_{cn} E_{Jn})}$, where $\Omega_n$ is the qubit's resonant frequency, and $E_{cn} = e^2/C_n$ and $E_{c\_nm} = e^2/2C_{nm}$. Therefore, using the Langevin equation [29], the system's equation of motion in a general form can be expressed as:

$$\dot{b}_n = -\left( j\Omega_n + \frac{\kappa_n}{2} \right) b_n + \frac{j}{\hbar} \frac{I_{bn}}{\sqrt{2\pi}} \left( \frac{8 E_{cn}}{E_{Jn}} \right)^{0.25} \frac{\Phi_0}{2\pi} + \frac{j}{\hbar} \frac{E_{cn}}{3} \left( b_n^+ + b_n \right)^3 + \frac{m_n}{\hbar} \frac{I_{bn}}{\sqrt{2}} \left( \frac{E_{Jn}}{8 E_{cn}} \right)^{0.25} (2e) V_{rf} + \sqrt{2\kappa_n} b_{n\_in} \\ - \sum_{\substack{m=1 \\ m \neq n}}^{4} \frac{j 4}{\hbar} E_{c\_nm} \left( \frac{E_{Jn}}{8 E_{cn}} \right)^{0.25} \left( \frac{E_{Jm}}{8 E_{cm}} \right)^{0.25} \frac{\left( b_m - b_m^+ \right)}{2} \tag{11}$$

where $\kappa_n$ and $b_{n\_in}$ is the qubit's decay rate and thermal noise arising due to the quantum circuit interaction with the environment. In the same way as the previously discussed systems, the Lindblad master equation is applied to govern the dynamics of the systems. The data listed in Table.3 are used to simulate and analyze this system.

Table 3. Data for simulation of 4-Qubits [18]

| | |
|---|---|
| $C_J$ (Josephson junction capacitance) | 6.24 pF |
| $C_{in}$ (input capacitance) | 0.08 pF |
| $C_{c1}$ (coupling capacitance ($Qu_1$ & $Qu_2$)) | 0.08 pF |
| $C_{c2}$ (coupling capacitance ($Qu_2$ & $Qu_3$)) | 0.08 pF |
| $C_{c3}$ (coupling capacitance ($Qu_3$ & $Qu_4$)) | 0.08 pF |
| $F_{c1}$ (charging energy frequency) | 606 MHz |
| $F_{J0}$ (Josephson energy frequency) | 5.2 GHz |
| $\kappa$ (Qubits' decay rate) | $0.022 \times F_{J0}$ |
| $V_{rf}$ (RF source voltage) | $1.5 \times 10^{-7}$ V |
| Tem (environment temperature) | 50 mK |

**InP HEMT coupling with the two external oscillators ($OS_I$_InP HEMT_$OS_{II}$)**

Finally, a cryogenic InP HEMT coupled with two external oscillators is discussed as the final quantum system. The schematic of the system is shown in Fig. 4. The system has been studied quantum mechanically, and its various quantum features have been investigated, including the InP HEMT nonlinearity effect on nonclassicality [21], the relationship between quantum correlation and noise figure [19], the InP HEMT nonlinearity effect on quantum discord and classical discord [20], and the generation of the quantum correlation at zero-IF band [22]. In this article, the Hamiltonian of the system is briefly discussed, and the Lindblad master equation is applied to solve the system dynamics. The Hamiltonian of the circuit has been theoretically derived, and the time evolution of the state of the coupled LCs has been investigated [19, 22]. The theory related to the system Lagrangian, classical Hamiltonian, and quantum Hamiltonian can be found in [19-22]. The quantum Hamiltonian of the system ($H = H_0 + H_{int}$), where $H_0$ is the free evolution, and $H_{int}$ is the interaction Hamiltonian, is defined as follows [19]:

$$H = \{\hbar\Delta_1(a_1^+ a_1) + \hbar\Delta_2(a_2^+ a_2) - \hbar\gamma_{q_1 q_2}(a_1 - a_1^+)(a_2 - a_2^+) - j\hbar\gamma_{q_1\varphi_2}(a_1 - a_1^+)(a_2 + a_2^+)$$
$$- j\hbar\gamma_{q_2\varphi_2}(a_2 - a_2^+)(a_2 + a_2^+) - \hbar\gamma_{q_1}(a_1 + a_1^+) + \hbar\gamma_{q_2}(a_2 + a_2^+) - j\hbar\gamma_{\varphi_1}(a_1 - a_1^+) - j\hbar\gamma_{\varphi_2}(a_2 - a_2^+)\}_L \quad (12)$$
$$+ \{-\hbar\gamma_{q_1 q_1\varphi_2}(a_1 - a_1^+)^2(a_2 + a_2^+) - \hbar\gamma_{q_2 q_2\varphi_2}(a_2 + a_2^+)(a_2 - a_2^+)^2 + j\hbar\gamma_{q_1\varphi_2\varphi_2}(a_1 - a_1^+)(a_2 + a_2^+)^2$$
$$+ \hbar\gamma_{\varphi_2\varphi_2\varphi_2}(a_2 + a_2^+)^3 + j\hbar\gamma_{q_2\varphi_2\varphi_2}(a_2 - a_2^+)(a_2 + a_2^+)^2 - j\hbar\gamma_{q_1\varphi_1\varphi_2}(a_1 - a_1^+)(a_1 + a_1^+)(a_2 + a_2^+)\}_{NL}$$

where $a_i$ and $a_i^+$ (i = 1, 2) stand for the annihilating and creating operators for the defined oscillators. The first two terms represent the free evolution related to the individual oscillator energy, and other terms determine the interaction Hamiltonian. The interaction Hamiltonian is divided into two parts, linear and nonlinear. All constants in Eq. 12 are given in detail in [29-22]. In addition, $\Delta_1$ and $\Delta_2$ are the external oscillators detuning frequencies. In the same way as the latter discussed systems, the Lindblad master equation is applied to govern the dynamics of the systems. The data listed in Table.4 are used to simulate and analyze this system.

Table 4. Data for the small signal model of the 4×50 μm InP HEMT at ~4.2 K [32-33].

| | |
|---|---|
| $R_g$ (Gate resistance) | 0.3 Ω |
| $L_g$ (Gate inductance) | 75 pH |
| $L_d$ (Drain inductance) | 70 pH |
| $C_{gs}$ (Gate-Source capacitance) | 107 fF |
| $C_{ds}$ (Drain-Source capacitance) | 51 fF |
| $C_{gd}$ (Gate-Drain capacitance) | 60 fF |
| $R_i$ (Gate-Source resistance) | 0.07 Ω |
| $R_j$ (Gate-Drain resistance) | 8 Ω |
| $g_d$ (Drain-Source conductance) | 12 mS |
| $g_m$ (Intrinsic transconductance) | 82 mS |
| $V_g$ (Gate bias voltage) | 0.03 V |
| $V_d$ (Drain bias voltage) | 0.06 V |
| Tem (Operational temperature) | 4.2 K |
| $T_d$ (Drain noise temperature) | 450 K |

**Results and Discussions**

The factors such as quantum correlation, signal-to-noise ratio (SNR), and fidelity of the quantum system are discussed to compare the efficiency of the four different quantum systems. The results of the study are listed as follows.

*Quantum correlation:*

Quantum correlation between cavity modes is a measure of the degree to which the quantum states of different cavity modes are correlated. In a quantum system with multiple cavity modes, the quantum state of one mode can be correlated with the quantum state of another mode, even if the two modes are spatially separated [5-7]. Correlation can arise due to various physical mechanisms, such as coupling between the systems or the interaction with a common environment. Quantum correlation between cavity modes is an essential property of quantum systems, as it may enable the creation of entangled states and can be exploited for various quantum information processing [1-7]. Therefore, understanding and controlling the quantum correlation between cavity modes is an important aspect of quantum information processing, and it has important implications for the development of practical quantum devices and technologies. In order to compare the mentioned quantum systems, the nature and strength of the quantum correlation between cavity modes can be studied and compared with each other. Systems with stronger and more robust correlations may be more suitable for quantum applications. In the simulations conducted in this study, the quantum discord ($Q_{discord}$) and classical discord ($C_{discord}$) were analyzed as significant quantifiers to evaluate the quantum correlation between modes, as shown in Fig. 5. One of the interesting results is that the quantum systems that use different cavities, such as OC_MR_MC, OC_PD_MC, and $OS_I$_InPHEMT_$OS_{II}$, show the mixing behavior, which arises due to the system's nonlinearity effects.

The quantum discord of each system has be compared, and it is observed that the 4-Qubits system operating at 50 mK shows stronger correlation between the first and second qubits. However, the quantum discord value for OC_MR_MC and OC_PD_MC at 50 mK is also satisfactory, while the InP HEMT coupling to external oscillators operating at 4.2 K has a low quantum discord value. Another important observation is the phase difference between $Q_{discord}$ and $C_{discord}$ for different systems. From Fig. 5c, it is shown that the InP HEMT system exhibits a phase difference between $Q_{discord}$ and $C_{discord}$, which suggests that this system may exhibit more complex and non-classical behavior, making it more interesting and potentially useful for specific quantum applications but more challenging to optimize and control. In contrast, there is a negligible phase difference between $Q_{discord}$ and $C_{discord}$ for OC_MR_MC and OC_PD_MC, as illustrated in Fig. 5a and Fig. 5b, respectively.

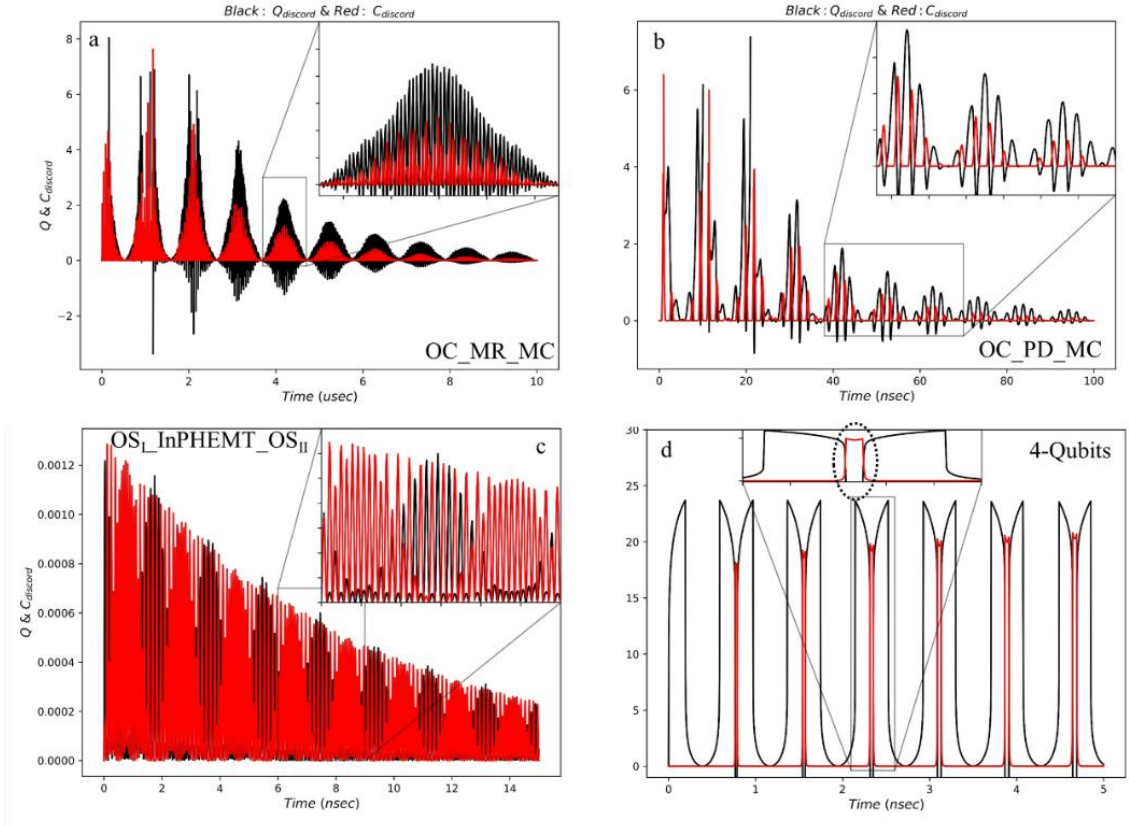

Fig. 5. Quantum discord and Classical discord as the quantifiers to compare the quantum and classical correlation between different modes; a) OC_MR_MC, b) OC_PD_MC, c) InP HEMT coupling with two external oscillators, d) 4-Qubits.

Another important point is the avoided level-crossing phenomenon observed in Fig. 5d, where the $Q_{discord}$ shows avoided level-crossing for special times, and the $C_{discord}$ is maximized. The avoided level-crossing phenomenon suggests that the quantum correlations between the qubits are changing in a non-trivial way, and this change is associated with a sudden increase in the classical correlations between the qubits. This phenomenon is well-known and studied in quantum mechanics, and it occurs when two energy levels of a quantum system approach each other and interact, causing a shift in the energy levels and a change in the system's behavior [23]. The results show that the avoided level-crossing is associated with a sudden increase in the classical correlations between the qubits, as measured by the $C_{discord}$. This suggests that the quantum correlations between the qubits are somehow being transformed into classical correlations, which could potentially be harnessed for quantum information and computation. The same phenomenon of transformation from quantum to classical correlation appears to occur in the InP HEMT system, although not to the same extent as the 4-Qubits system. If it were possible to drive the InP HEMT system around 50 mK, the avoided level-crossing phenomenon could arise, but not as strongly as in the 4-Qubits system. Nonetheless, the inconsistency in $Q_{discord}$ and $C_{discord}$ for the InP HEMT system, as compared to the other three systems, could because of the specific nature of the InP HEMT system, which operates at higher temperatures than that of the other three systems. This temperature difference could lead to different types and levels of noise and decoherence in the system, which could affect on the relationship between $Q_{discord}$ and $C_{discord}$.

*Signal-to-noise ratio (SNR)*

In addition to measuring quantum correlation, comparing the signal-to-noise ratio (SNR) values can help to determine which system is better suited for practical applications, where noise can significantly degrade the quantum signal quality. The SNR is a measuring parameter of the quality for the generated signal or measured by a quantum system relative to the level of the noise and unwanted effects exist in the system [8]. The SNR can be affected by various noise sources, such as thermal noise, electromagnetic interference, and measurement errors. Measuring the SNR of a quantum system is crucial because it provides information about the system's performance and suitability for different quantum applications. For example, in quantum communication or quantum sensing applications, a high SNR is necessary for reliable and accurate signal detection and processing [34-36]. Similarly, in quantum computing applications, a high SNR is important for maintaining the coherence of the quantum states and reducing errors in quantum operations [34].

A higher SNR generally indicates higher signal quality and lower noise level, leading to better performance and higher efficiency in various quantum applications. On the other hand, a lower SNR can result in increased errors, reduced fidelity, and lower overall performance of the quantum system. Therefore, measuring and comparing the SNR values of different quantum systems can help significantly to identify the most suitable system for specific quantum applications. The SNR results for the four different quantum systems are presented in Fig. 6 and compared with $Q_{discord}$ and $C_{discord}$. Surprisingly, there is a good consistency between the SNR and $C_{discord}$, rather than $Q_{discord}$. This suggests that $C_{discord}$, which is related to classical correlations between the subsystems, may be easier to measure and manipulate than $Q_{discord}$. Optimizing the $C_{discord}$ could therefore lead to better control and measurement of the system, improving the SNR. The figures also reveal that OC_MR_MC and OC_PD_MC, as well as InP HEMT, exhibit mixing behavior, affecting the cavity frequencies due to system nonlinearity. Fig. 6c differs from Fig. 6a and Fig. 6b, mainly due to the noise effect. The InP HEMT experiences more noise than the other two quantum systems. Finally, Fig. 6d investigates the impact of the interconnected capacitance on the SNR. The results show that the SNR between qubit 1 and qubit 2 with inter-connection capacitance ($C_{12}$) is higher than the SNR between qubit 2 and qubit 4 with $C_{24}$. This is due to the capacitance connection between the two mentioned qubits, where $C_{12} > C_{24}$.

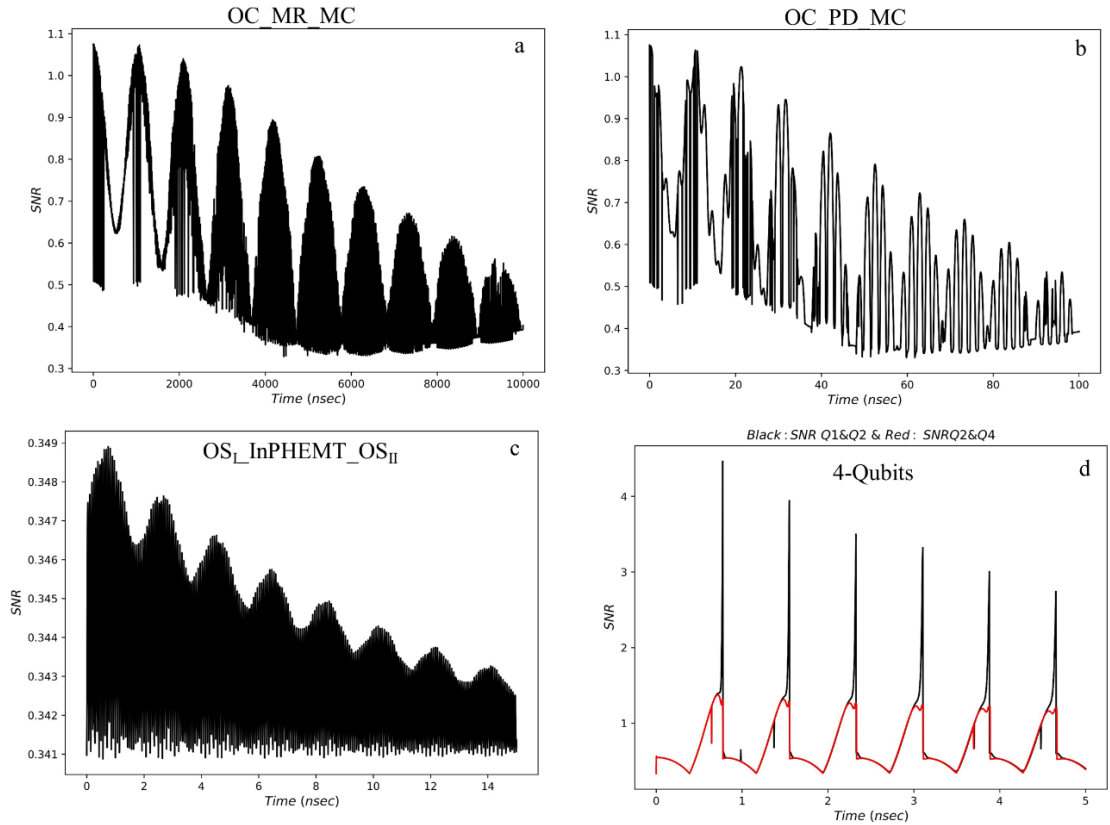

Fig. 6. SNR vs. time (nsec); a) OC_MR_MC, b) OC_PD_MC, c) InP HEMT coupling with two external oscillators, d) 4-Qubits.

*Fidelity*

The fidelity of a quantum system is a measure of how well it preserves the quantum state during operations such as measurement or manipulation. It compares the actual quantum state produced by the system with the target or ideal quantum state [37-39]. Various noise sources and errors, such as decoherence and measurement errors, can affect fidelity in a quantum system. Measuring fidelity is important as it provides information about the system's performance and suitability for different quantum applications. For instance, high fidelity is crucial for reliable and accurate quantum operations in quantum computing [37-38], while in quantum communication and quantum sensing, high fidelity is essential for achieving accurate and reliable signal detection and processing [39]. A higher fidelity implies higher accuracy and precision, leading to better performance and higher efficiency in various quantum applications. The fidelity of the four different quantum systems OC_MR_MC, OC_PD_MC, InP HEMT coupling to external oscillators, and 4-Qubits is calculated versus α, which is the coherent coefficient, and illustrated in Fig. 7. Similarly, to the SNR results, OC_MR_MC and OC_PD_MC behave in the same way, while InP HEMT and 4-Qubits show similar behavior. This similarity may be due to shared common features or underlying mechanisms that affect their fidelity similarly as α is varied. For example, how α affects the energy levels and transition probabilities in the different systems could be similar, leading to similar fidelity behavior. The observed similarity in fidelity behavior between InP HEMT and 4-Qubits has interesting practical implications for designing and optimizing quantum systems for specific applications. For example, InP HEMT is suitable for use in the interface electronics of quantum computing. Therefore, by understanding the common factors that effect on fidelity in these systems, researchers may identify general principles or techniques for improving fidelity in a broad range of quantum systems.

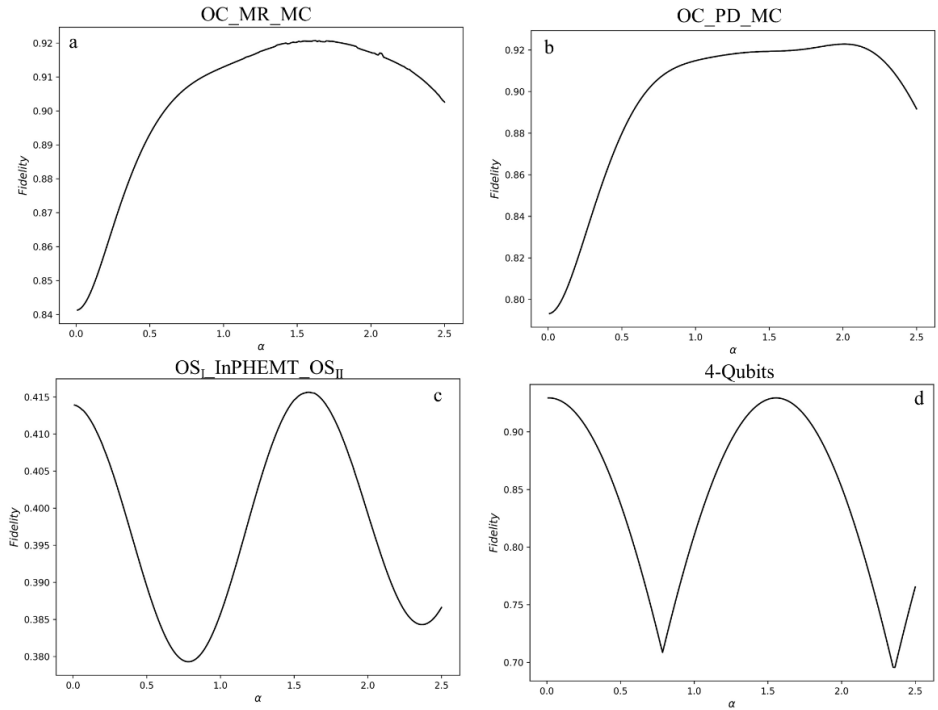

Fig. 7 Fidelity vs α; a) OC_MR_MC, b) OC_PD_MC, c) InP HEMT coupling with two external oscillators, d) 4-Qubits.

*Scalability*

Another essential factor to consider is the scalability of the system. A quantum system that is easily scalable may be more practical for future applications [40]. Although the scalability of the four systems was not explicitly evaluated, the number of qubits and cavities, interconnectivity, control and measurement, and materials and fabrication were identified as the significant factors to consider in assessing the scalability of a quantum system. Considering these factors, one can determine which of the four quantum systems is more efficient for quantum applications.

Consequently, by comparing the simulation results of the quantum correlation, SNR, and fidelity, it may be possible figure out the relationships between the mentioned quantities in a quantum system. The relationships may be complex and dependent on various factors, such as the specific system, the measurement technique, the type and level of noise in the system. In general, increasing the quantum correlation between modes of a quantum system can lead to increase in the SNR of the system since the correlation can enhance the coherent signal between the modes while reducing the effect of noise. This can result in a higher-quality signal and a better SNR. In the same way, increasing the quantum correlation can also potentially enhance the fidelity of a quantum system, since the correlation can help to maintain the coherence and stability of the quantum states, and therefore leading to higher accuracy and precision in quantum operations and measurements.

**Conclusions:**

In this article, four different kinds of quantum systems have been studied similarly, namely OC_MR_MC, OC_PD_MC, 4-Qubits, and two external oscillators coupled through InP HEMT. To study which quantum system is suitable for quantum applications, some important properties such as the quantum correlation between cavity modes, the signal-to-noise ratio (SNR), and the fidelity of each system have been investigated. The simulation results show that the quantum correlation and fidelity are affected by various factors, such as the coherent coefficient α and the system's temperature. We have also observed interesting phenomena in some systems, such as avoided level crossing and partial in-phase correlation. It was found that the SNR and $C_{discord}$ have good consistency, while $Q_{discord}$ does not show the same level of consistency. This finding suggests that $C_{discord}$ may be a more useful measurement of quantum correlation than $Q_{discord}$ in some instances. In addition, it was observed a similarity in the behavior of fidelity versus α for OC_MR_MC and OC_PD_MC on the one hand, and for InP HEMT coupling to external oscillators and 4-Qubits on the other hand. This similarity could indicate that the systems share some common features or underlying mechanisms that similarly affect their fidelity. Finally, the OC_MR_MC and OC_PD_MC, and 4-Qubits systems appear more reliable and efficient for quantum applications due to their consistent behavior and stable performance. However, the InP HEMT system could offer unique advantages and opportunities for certain quantum technologies.

Conflict of interest: There is no conflict of interest for this work.

Ethics approval and consent to participate: I confirm that this work is original and has been neither published nor is currently under consideration for publication elsewhere.

Consent for publication: The author of this study gives the publisher the permission of the author to publish the work.

Availability of data and materials: There are no datasets generated for this work.

Competing interests: There are no competing interests.

Funding: There is no funding for this work.

Authors' contributions: All of the studies have been done by Ahmad Salmanogli and Vahid Sharif Sirat.